\documentclass{article}
\usepackage{spconf,amsmath,graphicx,hyperref}
\usepackage{xcolor}
\usepackage{multirow}

\title{Probing Whisper for Dysarthric Speech in Detection and Assessment}
\name{Zhengjun Yue$^{1}$, Devendra Kayande$^{1}$, Zoran Cvetkovic$^{2}$, and Erfan Loweimi$^{3}$}

\address{
$^{1}$TU Delft, the Netherlands; $^{2}$King's College London, UK; $^{3}$Cisco, UK \\
}

\begin{document}
%
\maketitle
\begin{abstract}
Large-scale end-to-end models such as Whisper have shown strong performance on diverse speech tasks, but their internal behavior on pathological speech remains poorly understood. Understanding how dysarthric speech is represented across layers is critical for building reliable and explainable clinical assessment tools. This study probes the Whisper-Medium model encoder for dysarthric speech for detection and assessment (i.e., severity classification). We evaluate layer-wise embeddings with a linear classifier under both single-task and multi-task settings, and complement these results with Silhouette scores and mutual information to provide perspectives on layer informativeness. To examine adaptability, we repeat the analysis after fine-tuning Whisper on a dysarthric speech recognition task. Across metrics, the mid-level encoder layers (13–15) emerge as most informative, while fine-tuning induces only modest changes.
The findings improve the interpretability of Whisper’s embeddings and highlight the potential of probing analyses to guide the use of large-scale pretrained models for pathological speech.
\end{abstract}

\section{Introduction}

Dysarthria is a speech disorder resulting from neurological impairments that affect articulation, phonation, and prosody \cite{kent2000research,theodoros1994perceptual,darley1969clusters}. Accurate detection and assessment of the severity level of dysarthric speech are essential for early diagnosis, therapeutic intervention, and assistive speech technologies \cite{kodrasi2020spectro,sekhar2022dysarthric,al2024detection,sajiha2024automatic}. Traditional methods relied on handcrafted features or statistical models, but recent advances in deep learning and large-scale models have enabled more robust and generalizable representations for dysarthric speech \cite{javanmardi2024exploring,hu2024self}.

OpenAI’s Whisper \cite{radford2023robust}, as a large-scale weakly supervised model, trained on diverse multilingual and multitask data, has demonstrated promising performance in various speech-related tasks \cite{liu2024sparsely,lyu2024real}. While primarily designed for automatic speech recognition (ASR), its encoder layers capture hierarchical speech representations that can benefit various downstream tasks \cite{zezario2024study,koenecke2024careless,zhang2024whisper}, where selecting the optimal layer for downstream tasks is critical.
Recent work has begun to investigate Whisper's internal embeddings for pathological speech analysis \cite{patel2024noise,vinotha2024leveraging,charlesworth2024automatic,agaoglu2024does}. However, the effectiveness of different layers for dysarthric detection and assessment, as well as how these representations are influenced by fine-tuning \cite{mirella2024improving}, remains insufficiently understood.

This gap motivates our study to systematically probe Whisper embeddings and determine the most informative layer for dysarthric speech detection and assessment, and improve the interpretability of the large-scale models in clinical applications. 
Specifically, we address three research questions: 1) How does multi-task learning compare to single-task learning for dysarthric speech detection and assessment? 2) Which Whisper embedding layers are most informative for each task? and 3) How does fine-tuning alter the representations across layers and affect classification performance? 

To answer these questions, we conduct a comprehensive layer-wise probing analysis of Whisper-medium (Whisper-M) embeddings. We evaluate layer representations using multiple criteria: (i) performance (accuracy and F1-score) across single-task and multi-task settings to evaluate their relative effectiveness, (ii) mutual information (MI) \cite{scikit-learn} between the embeddings and dysarthria labels to measure feature relevance, and (iii) Silhouette scores \cite{scikit-learn} to assess the clustering quality of dysarthric and non-dysarthric speech and severity levels. 
To study adaptability, we further compute the MI between pretrained (PT) and fine-tuned (FT) embeddings to examine how fine-tuning affects different layers.

\section{Methodology}

\begin{figure}[t]
\centering
\includegraphics[width=0.85\columnwidth]{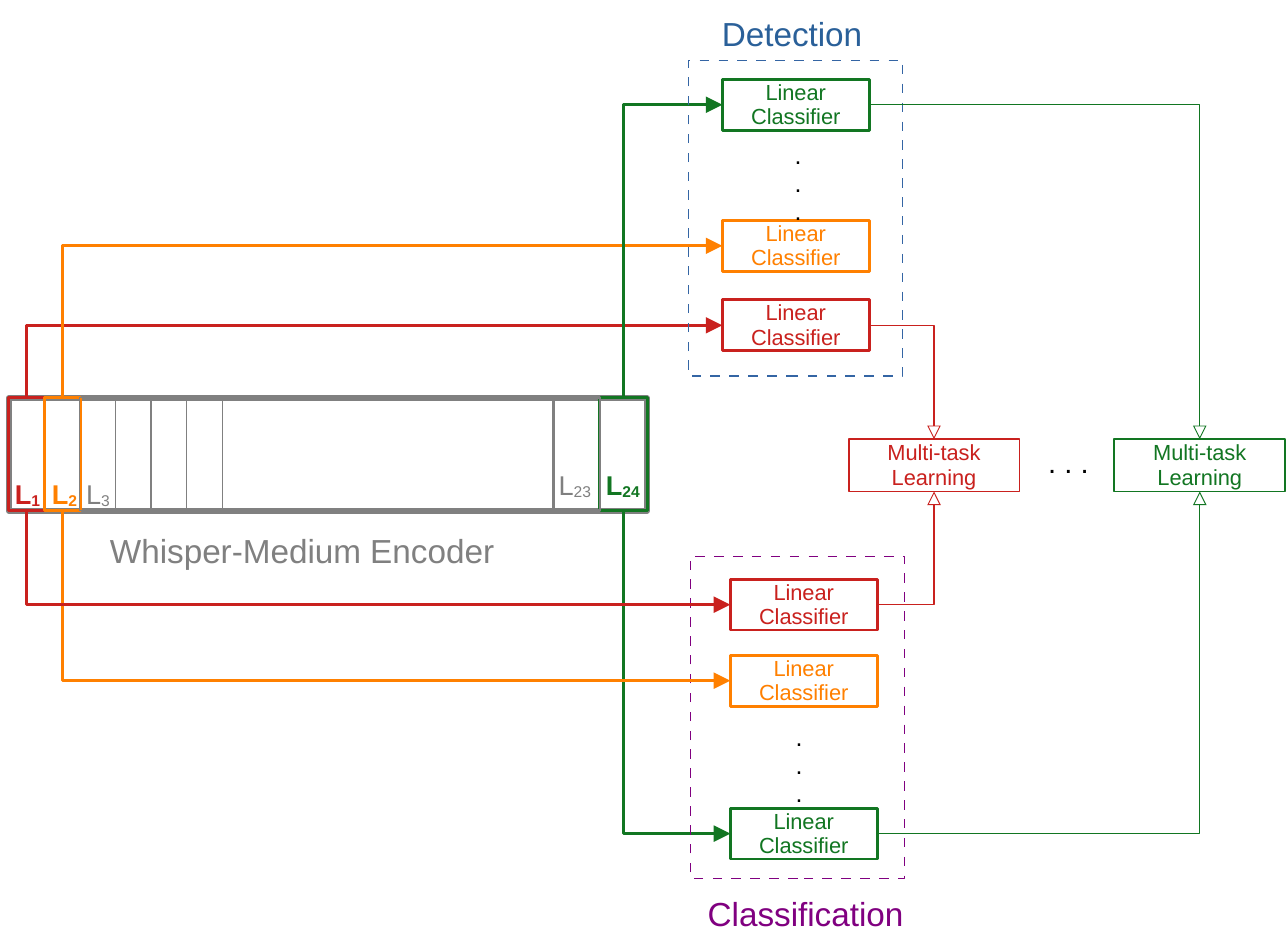}
\vspace{-2mm}
    \caption{Workflow of layer probing on Whisper-M encoder layers' embeddings using the single and multi-task approaches.}  
    \label{fig:workflow}
\end{figure}

The Whisper-M encoder, taking 80-D FBank as inputs, consists of 24 layers that generate hierarchical speech representations. To investigate the effectiveness of these embeddings and to determine the most informative layer in the Whisper-M encoder for dysarthric speech detection and classification, we extract representations from all layers. Fig.~\ref{fig:workflow} shows the workflow for single- and multi-task learning. 

\vspace{-2mm}
\subsection{Single- and multi-task detection and assessment}

For the \textbf{single-task learning} setup, we train separate linear classifiers (a fully-connected layer) for each layer to independently perform either detection (dysarthric vs. non-dysarthric speech) or assessment (identifying dysarthria severity levels).

In the \textbf{multi-task learning} setup, we introduce a shared learning framework where a classifier jointly predicts detection and severity from the same embeddings. The goal is to leverage shared speech representations across tasks, potentially improving itself or the other task's performance by enabling joint learning of relevant features. 
The objective function is defined as the sum of the detection and severity classification losses, encouraging the model to learn features that benefit both tasks. 
By comparing the single- and multi-task setups across different layers, we aim to determine the most effective learning strategy and the optimal layer for dysarthric speech detection and assessment.

\subsection{Probing analysis and evaluation metrics}

We conduct a layer-wise probing analysis using three evaluation techniques: detection and severity classification accuracy, Mutual Information (MI), and Silhouette score. Each metric captures a different perspective on how well embeddings encode dysarthria-related information, helping to identify the optimal layer for downstream tasks.

\subsubsection{Detection and assessment performance}
For each layer, we evaluate task performance using both accuracy and F1-score.
Accuracy reflects the overall correctness of predictions, while the F1-score provides a harmonic mean of the precision and recall, which is particularly important given the class imbalance in dysarthric speech datasets.
Higher values on these metrics indicate that the corresponding layer captures more discriminative representations.

\subsubsection{Mutual information}

Unlike accuracy and F1-score, which assess performance through a classifier, the MI score quantifies the intrinsic dependency between embeddings and dysarthria labels. Higher MI values indicate that the embeddings inherently encode more label-relevant information, independent of classifier training. We compute MI for each layer to assess how well the Whisper encoder layers capture dysarthria-related patterns. Layers with high MI are therefore expected to be more informative for detection or assessment. 

We also compute MI between pretrained and fine-tuned embeddings to examine how fine-tuning alters the layer representations (e.g., which layers are most affected) and influences classification performance. Lower MI values suggest that FT significantly modifies the embeddings, while higher values indicate that the original representation structure is largely preserved after fine-tuning.

\subsubsection{Silhouette score}
The silhouette score evaluates the clustering quality of the embeddings by measuring how well embeddings of dysarthric and non-dysarthric speech, or different severity levels, are separated in the corresponding space. This score ranges from -1 to 1, with higher values indicating more distinct and well-separated clusters. In our setting, high Silhouette scores imply that a layer produces embeddings that naturally distinguish dysarthria-related categories.

\section{Experimental setup}
\label{sec:exp}

\subsection{Dataset}

We use TORGO \cite{rudzicz2012torgo}, a widely used English dysarthric dataset, which contains 21 hours of speech collected from 15 speakers: 8 dysarthric speakers with different severity levels (\textit{severe}, \textit{moderate to severe \footnote{We merge this severity level with severe in the assessment task}}, \textit{moderate}, and \textit{mild}, totaling 7.3 hours) and 7 typical speakers (13.7 hours). 
The dataset consists of both isolated word (615 unique items) and sentence utterances (354 unique items), with a total vocabulary size of 1573. This diversity makes TORGO well-suited for evaluating dysarthria detection and assessment methods.

\begin{table}[t]
\centering
\caption{Utterance numbers of training and test set split for different speaker groups (fold 1). The numbers of utterances are the same for folds 1, 2 and 3 and in fold 4 and fold 5 it is +1 or -1 for each class without a major difference.}
\vspace{2mm}
\resizebox{0.9\linewidth}{!}{ 
\begin{tabular}{l|ccc|cc}
\hline
\textbf{Split} & \textbf{Mild} & \textbf{Moderate} & \textbf{Severe} & \textbf{Dys} & \textbf{Typ} \\ 
\hline
\textbf{Train} & 1228  & 1138  & 2544  & 4910  & 9158  \\
\textbf{Test}  & 307   & 285   & 636   & 1228  & 2290  \\
\hline
\end{tabular}
}
\label{tab:train-test-split}
\end{table}

\subsection{Feature extraction}

As baselines, we extracted filterbank (\textbf{FBank}) features using \texttt{torchaudio} \cite{yang2022torchaudio}. Four sets of FBank features were computed: 80- and 128-D, and the same features concatenated with three pitch-related features \cite{ghahremani2014pitch} extracted with \texttt{librosa} \cite{mcfee2015librosa}, yielding 83- and 131-D features.

We extract embeddings from the Whisper-M model \cite{radford2023robust},  a Transformer \cite{vaswani2017attention} with 24 encoder and 6 decoder layers. Each encoder layer outputs a 1024-D embedding. For every utterance in TORGO, we extracted embeddings from all 24 encoder layers and averaged them across time.

To analyze the effect of domain adaptation, we fine-tuned \footnote{With 3000 steps, learning rate of 1e-5 and batch size of 8.} 
Whisper-M on the TORGO dataset for the ASR task, following the split in \cite{yue2020autoencoder}. Then, embeddings were extracted from FT models for probing.

\subsection{Classifier and training setup}
To analyze the raw performance of each layer, we employed a simple linear classifier consisting of a single fully-connected layer with softmax activation. For dysarthria detection, the classifier head has two units (1024$\times$2 parameters), where 1024 is the embedding size. For severity classification, the classifier head has four units (1024$\times$4) in the single-task setup. In the multi-task setup, the classifier head branched into two outputs: one with two units for detection and another with four units for severity classification. 
We used Stratified-KFold cross validation (K=5) \cite{Kohavi1995,scikit-learn}, to ensure balanced class distributions in training and test splits (Table~\ref{tab:train-test-split}). Each classifier was trained for 20 epochs per fold using the AdamW optimizer \cite{zhou2024towards} with a learning rate of 3e-4 and a batch size of 32. Cross-entropy loss was used in both single-task and multi-task setups; in the latter, losses from the detection and severity heads were summed with the same weights. 
Performance was reported as the average results over the five folds, along with the standard deviation, visualized with error bars. The same training setup was applied to classifiers trained on embeddings extracted from each Whisper layer (both PT and FT) as well as the FBank features.

\section{Result and Discussion}

\subsection{Single- vs multi-task learning}

\begin{figure}[t]
\centering
\includegraphics[width=0.9\columnwidth]{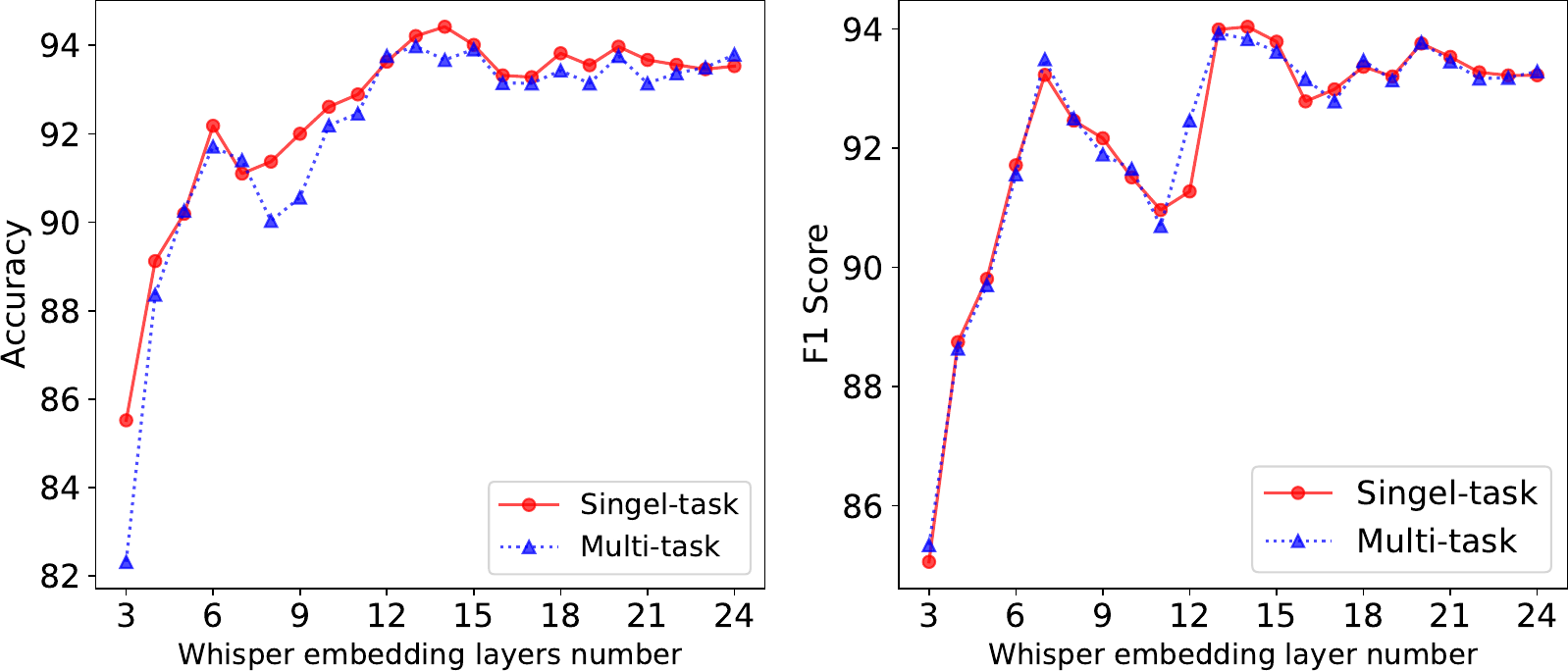}
\vspace{-1mm}
    \caption{Detection results for probing pretrained Whisper embeddings (single-task vs. multi-task).}  
    \label{fig:det-mtl}
\end{figure}

Figs.~\ref{fig:det-mtl} show the accuracy and F1-score of the detection task under single-task and multi-task setups\footnote{Severity classification figures show the same trends and were omitted due to space constraints.}. 
Contrary to expectations of a regularization effect, multi-task learning provides minimal benefit over single-task learning across both metrics. This is likely because the two tasks are highly related: in the detection task, all dysarthric classes are treated as a single category, while the severity classification task subdivides them into severity levels. As a result, multitask learning does not introduce significant additional information.

Probing results illustrate that the layers 13--15 yield the best performance for both single- and multi-task setups. While earlier layers capture lower-level acoustic information and later layers specialized for ASR, these middle layers appear to balance task-specific and general representations most effectively. An interesting observation is that performance does not decline significantly after the optimal layers (16--24). This suggests that the embeddings from higher layers, particularly the final encoder layer\textemdash despite being primarily tailored for speech recognition\textemdash remain highly effective in distinguishing between dysarthric and typical speech. This implies that the Whisper encoder does not map all speech types into a single canonical representation but instead projects them into distinct subspaces optimized for ASR. These separate subspaces ensure that even the embeddings from the highest encoder layer remain useful for both detection and classification tasks.

\begin{figure}[t]
\centering
\includegraphics[width=0.6\columnwidth]{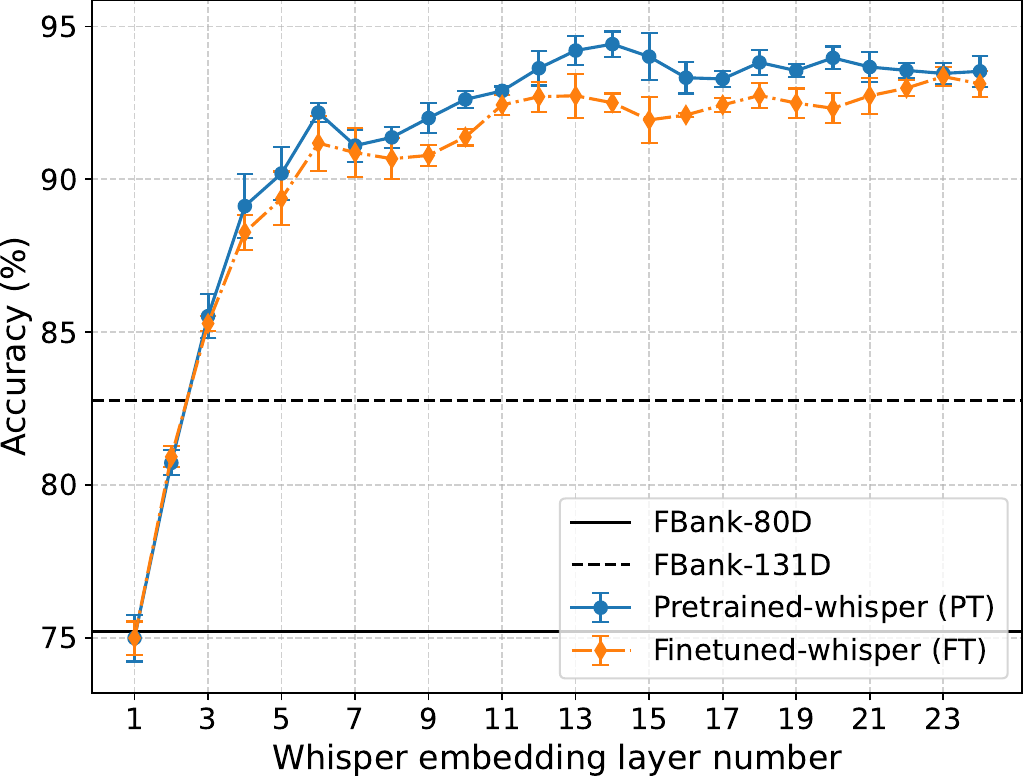}
\vspace{-2mm}
    \caption{Detection accuracy error bar for probing pretrained and finetuned Whisper embeddings.}  
    \label{fig:det-errbar}
\end{figure}

\subsection{Fine-tuning effect}

Fig.~\ref{fig:det-errbar} show detection performance after fine-tuning Whisper on the TORGO ASR task, along with error bars across five folds and baselines using 80- and 131-D FBank features. As expected, the performance of the earliest Whisper layers is comparable to FBank-80, which matches the model’s training setup as the Whisper-M model is trained with FBank-80 features. Adding pitch features to higher-dimensional FBank representations (Fbank-131) improves performance to a level similar to the 3rd Whisper layer. The small error bars indicate consistent results across folds.

Another observation is that fine-tuning has minimal impact on the performance of classifiers built on the earliest and latest layers, albeit for different reasons. In the lower layers, weak gradients lead to minimal weight updates, so the embeddings change very little. In the higher layers, although gradients are stronger, the Whisper model already maps diverse speech types into well-separated subspaces. These subspaces remain effective for linear classifiers, yielding high performance even though the embeddings were primarily optimized for ASR rather than for detection or severity classification.

Table~\ref{tab:stl-mtl} summarizes the FT and PT results, reporting the performance of the best layers (based on probing) on the detection and severity assessment tasks, together with the FBank feature baseline results.

\begin{table}[t]
\centering
\renewcommand{\arraystretch}{1.2}
\setlength{\tabcolsep}{10pt}
\caption{Detection and classification accuracy for different feature sets and tasks. 
``*'' indicates the performance of the embeddings from the optimal layer of Whisper-M. ST: Single-task, MT: multi-task.}
\vspace{2mm}
\resizebox{\linewidth}{!}{ 
\begin{tabular}{l|cc|cc}
\hline
\textbf{Feature Set} & \multicolumn{2}{c|}{\textbf{Detection}} & \multicolumn{2}{c}{\textbf{Assessment}} \\ 
\cline{2-5} 
& \textbf{ST} & \textbf{MT} & \textbf{ST} & \textbf{MT} \\ 
\hline
\textbf{Fbanks-80/83}   & 75.2/78.0 & 74.9/77.8 & 72.2/73.8 & 71.7/73.8 \\ 
\textbf{Fbanks-128/131} & 81.2/82.8 & 80.7/81.1 & 78.4/80.1 & 78.3/79.5 \\ 
\textbf{Whisper-PT*}    & \textbf{94.4} & \textbf{94.0} & \textbf{94.1} & \textbf{93.7} \\ 
\textbf{Whisper-FT*}    & 93.4 & 93.4 & 93.5 & 93.2 \\ 
\hline
\end{tabular}
}
\label{tab:stl-mtl}

\end{table}

\subsection{Mutual information and Silhouette score analysis}
\begin{figure}[t]
\centering
\includegraphics[width=\columnwidth]{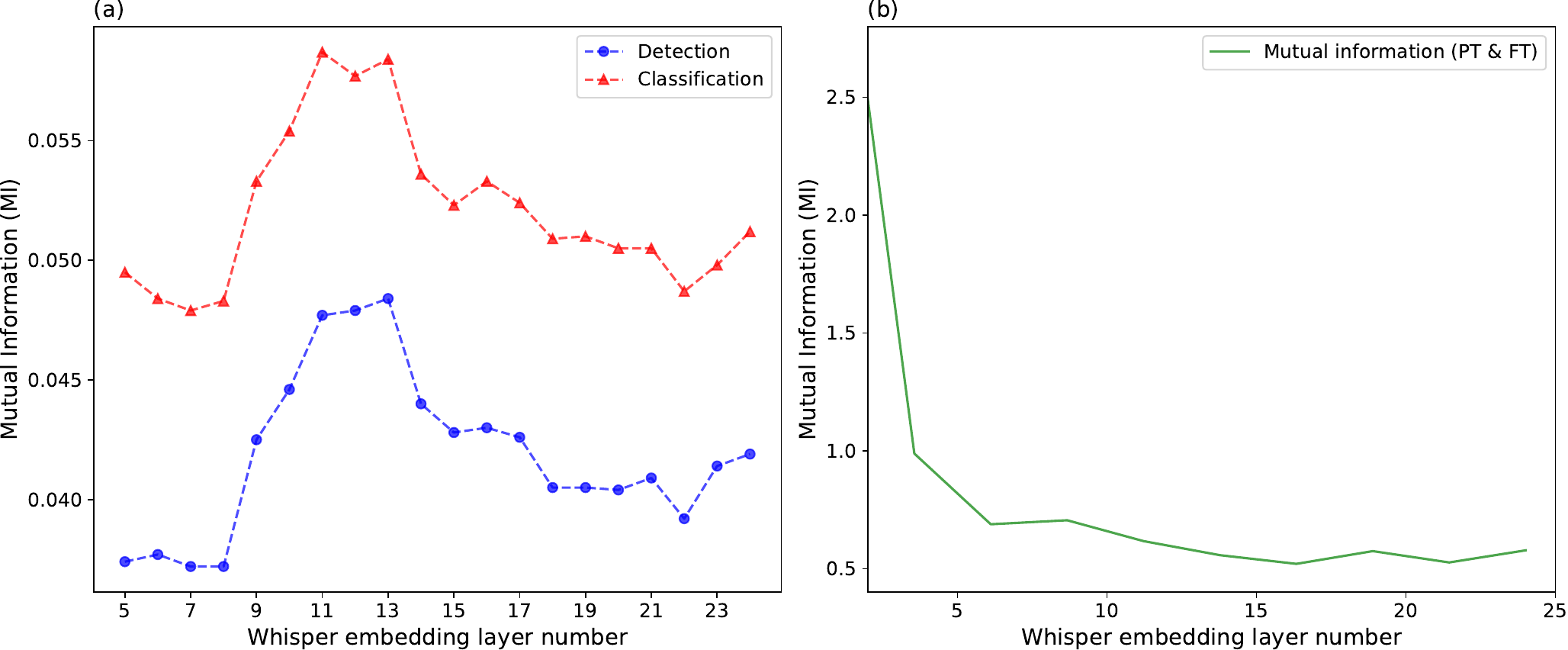}
\vspace{-5mm}
    \caption{Layer-wise MI between (a) Whisper PT embeddings and labels, (b) PT and FT whisper embeddings.}  
    \label{fig:det-class-MI-pt}
\end{figure}

While probing with linear classifier \cite{raymondaud2024probing} evaluates task performance directly, MI provides a task-agnostic view of representational relevance. 
Fig.~\ref{fig:det-class-MI-pt} shows the MI between embeddings of each layer and dysarthria labels in the detection and severity classification tasks. Results peak at layer 13, consistent with probing results for the task performance
(Figs.~\ref{fig:det-mtl}) 
that identified layers 13-15 as optimal. Unlike classifier-based probing, this finding is obtained without training.
We also calculated the MI between embeddings of the PT and FT Whisper models. This analysis, which is task-agnostic as it does not rely on labels, quantifies how MI varies across different layers after FT. The results show that MI values decrease progressively from the lower to the higher layers. This trend can be perfectly explained by the fact that the backpropagated gradient is stronger in the higher layers during FT, leading to more substantial modifications, whereas the lower layers receive a weaker gradient and thus undergo minimal changes, resulting in higher MI values.

\begin{figure}[t]
\centering
\includegraphics[width=0.65\columnwidth]{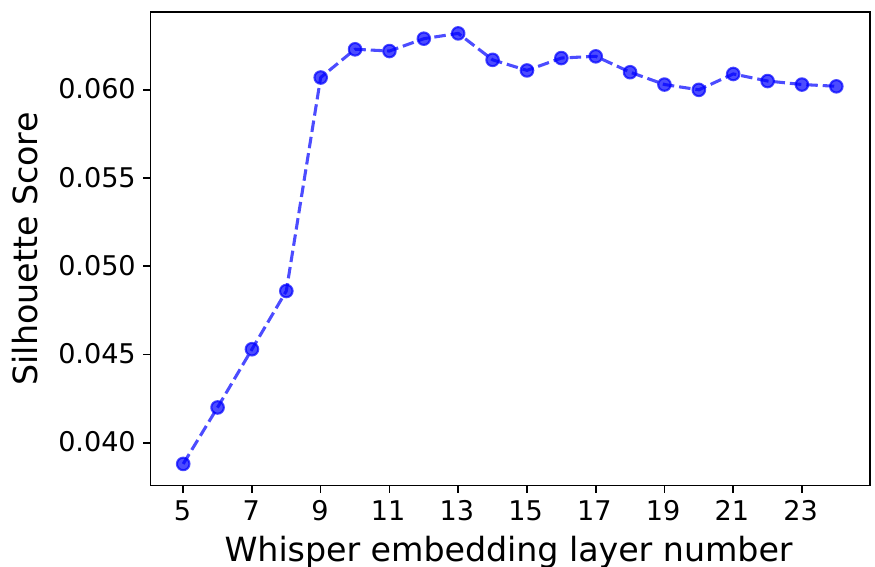}
\vspace{-2mm}
    \caption{Silhouette score between layer-wise PT whisper embeddings and labels for the detection task.}  
    \label{fig:det-Sil-pt}
\end{figure}

Finally, we probed the Whisper encoder layers using the Silhouette Score metric, which measures how well-separated different class clusters are in the feature space. Fig.~\ref{fig:det-Sil-pt} presents the results for the detection task\footnote{For severity classification, they are less meaningful as the metric assumes well-separated clusters whereas severity is subjective and overlapping.}. 
As shown, the Silhouette Score peaks around layer 13, aligning closely with both accuracy, F1-score and MI analyses.

This consistency is particularly intriguing from a technical perspective, as these metrics stem from fundamentally different modelling approaches. Silhouette score is a geometric clustering measure, mutual information quantifies statistical dependence between representations and class labels, and classification accuracy/F1-score directly evaluates task performance using a trained classifier. The fact that all metrics peak around the same layer suggests that the representations learned by Whisper strike an optimal balance between structure and task-relevant information at this depth. Moreover, the minimal decline in Silhouette Score at higher layers further supports our hypothesis that the Whisper encoder maps different speech types into distinct subspaces rather than collapsing them into a single canonical representation.

\vspace{1mm}

\section{Conclusion}
We probed Whisper-M’s encoder layers for dysarthric speech detection and severity classification, finding that mid-level layers (13--15) consistently provided optimal performance across accuracy, F1-score, mutual information, and Silhouette Score. Multitask learning did not improve results. 
Fine-tuning on dysarthric ASR modified higher-layer representations but had limited impact on downstream classification, suggesting that pretrained Whisper embeddings already provide strong task-relevant features. 
Our findings offer two key insights: first, large-scale models trained for ASR implicitly encode clinically relevant information that can be repurposed for pathological speech assessment; second, probing across complementary metrics reveals consistent representational patterns, strengthening the reliability of layer selection strategies.
Future work includes applying other fundamental models and more diverse pathological speech datasets.

\vfill\pagebreak

\clearpage

\bibliographystyle{IEEEbib}
\bibliography{strings}

\begin{thebibliography}{10}

\bibitem{kent2000research}
R.~Kent,
\newblock ``Research on speech motor control and its disorders: A review and prospective,''
\newblock {\em Journal of Communication disorders}, vol. 33, no. 5, pp. 391--428, 2000.

\bibitem{theodoros1994perceptual}
D.~Theodoros, B.~Murdoch, and H.~Chenery,
\newblock ``Perceptual speech characteristics of dysarthric speakers following severe closed head injury,''
\newblock {\em Brain injury}, vol. 8, no. 2, pp. 101--124, 1994.

\bibitem{darley1969clusters}
F.~Darley, A.~Aronson, and J.~Brown,
\newblock ``Clusters of deviant speech dimensions in the dysarthrias,''
\newblock {\em Journal of speech and hearing research}, pp. 462--496, 1969.

\bibitem{kodrasi2020spectro}
I.~Kodrasi and H.~Bourlard,
\newblock ``Spectro-temporal sparsity characterization for dysarthric speech detection,''
\newblock {\em IEEE/ACM Transactions on Audio, Speech, and Language Processing}, vol. 28, pp. 1210--1222, 2020.

\bibitem{sekhar2022dysarthric}
S.~Sekhar, G.~Kashyap, A.~Bhansali, K.~Singh, et~al.,
\newblock ``Dysarthric-speech detection using transfer learning with convolutional neural networks,''
\newblock {\em ICT Express 2022}.

\bibitem{al2024detection}
A.~Al-Ali et~al.,
\newblock ``The detection of dysarthria severity levels using ai models: A review,''
\newblock {\em IEEE Access}, 2024.

\bibitem{sajiha2024automatic}
S.~Sajiha et~al.,
\newblock ``Automatic dysarthria detection and severity level assessment using cwt-layered cnn model,''
\newblock {\em EURASIP Journal on Audio, Speech, and Music Processing}, vol. 2024, no. 1, pp. 33, 2024.

\bibitem{javanmardi2024exploring}
F.~Javanmardi, S.~Kadiri, and P.~Alku,
\newblock ``Exploring the impact of fine-tuning the wav2vec2 model in database-independent detection of dysarthric speech,''
\newblock {\em IEEE journal of biomedical and health informatics}, 2024.

\bibitem{hu2024self}
S.~Hu et~al.,
\newblock ``Self-supervised asr models and features for dysarthric and elderly speech recognition,''
\newblock {\em IEEE/ACM TASLP 2024}.

\bibitem{radford2023robust}
J.~A.~Radford, A.and~Kim, T.~Xu, G.~Brockman, C.~McLeavey, and I.~Sutskever,
\newblock ``Robust speech recognition via large-scale weak supervision,''
\newblock in {\em International conference on machine learning}. PMLR 2023.

\bibitem{liu2024sparsely}
W.~Liu, Y.~Qin, Z.~Peng, and T.~Lee,
\newblock ``Sparsely shared lora on whisper for child speech recognition,''
\newblock in {\em IEEE ICASSP 2024}, pp. 11751--11755.

\bibitem{lyu2024real}
K.~Lyu, R.~Lyu, and H.~Chang,
\newblock ``Real-time multilingual speech recognition and speaker diarization system based on whisper segmentation,''
\newblock {\em PeerJ Computer Science}, vol. 10, pp. e1973, 2024.

\bibitem{zezario2024study}
R.~Zezario, Y.~Chen, S.~Fu, Y.~Tsao, H.~Wang, and C.~Fuh,
\newblock ``A study on incorporating whisper for robust speech assessment,''
\newblock in {\em IEEE ICME 2024}, pp. 1--6.

\bibitem{koenecke2024careless}
A.~Koenecke, A.~Choi, K.~Mei, H.~Schellmann, and M.~Sloane,
\newblock ``Careless whisper: Speech-to-text hallucination harms,''
\newblock in {\em The ACM Conference on Fairness, Accountability, and Transparency}, 2024, pp. 1672--1681.

\bibitem{zhang2024whisper}
L.~Zhang, N.~Jiang, Q.~Wang, Y.~Li, Q.~Lu, and L.~Xie,
\newblock ``Whisper-sv: Adapting whisper for low-data-resource speaker verification,''
\newblock {\em Speech Communication}, 2024.

\bibitem{patel2024noise}
H.~Patel and H.~Patil,
\newblock ``Noise robust whisper features for dysarthric automatic speech recognition,''
\newblock {\em Small}, vol. 12, no. 768, pp. 12, 2024.

\bibitem{vinotha2024leveraging}
R.~Vinotha, D.~Hepsiba, and LD. Anand,
\newblock ``Leveraging openai whisper model to improve speech recognition for dysarthric individuals,''
\newblock in {\em APCIT}. IEEE, 2024, pp. 1--5.

\bibitem{charlesworth2024automatic}
C.~Charlesworth,
\newblock ``Automatic dysarthria severity assessment using whisper-extracted features,''
\newblock 2024.

\bibitem{agaoglu2024does}
O.~Agaoglu,
\newblock ``How does openai’s whisper interpret dysarthric speech?,'' 2024.

\bibitem{mirella2024improving}
G~Mirella,
\newblock ``Improving state-of-the-art asr systems for speakers with dysarthria,'' 2024.

\bibitem{scikit-learn}
F.~Pedregosa et~al.,
\newblock ``{Scikit-learn: },'' 2011.

\bibitem{rudzicz2012torgo}
F.~Rudzicz, A.~Namasivayam, and T.~Wolff,
\newblock ``The torgo database of acoustic and articulatory speech from speakers with dysarthria,''
\newblock {\em Language Resources and Evaluation}, vol. 46, no. 4, pp. 523--541, 2012.

\bibitem{yang2022torchaudio}
Y.~Yang et~al.,
\newblock ``Torchaudio: Building blocks for audio and speech processing,''
\newblock in {\em IEEE ICASSP 2022}.

\bibitem{ghahremani2014pitch}
P.~Ghahremani et~al.,
\newblock ``A pitch extraction algorithm tuned for automatic speech recognition,''
\newblock in {\em IEEE ICASSP 2014}, pp. 2494--2498.

\bibitem{mcfee2015librosa}
B.~McFee et~al.,
\newblock ``librosa: Audio and music signal analysis in python.,''
\newblock in {\em SciPy}, 2015, pp. 18--24.

\bibitem{vaswani2017attention}
A~Vaswani,
\newblock ``Attention is all you need,''
\newblock {\em Advances in Neural Information Processing Systems}, 2017.

\bibitem{yue2020autoencoder}
Z~Yue et~al.,
\newblock ``Autoencoder bottleneck features with multi-task optimisation for improved continuous dysarthric speech recognition,''
\newblock in {\em Interspeech 2020}.

\bibitem{Kohavi1995}
Ron K.,
\newblock ``A study of cross-validation and bootstrap for accuracy estimation and model selection,''
\newblock {\em IJCAI 1995}.

\bibitem{zhou2024towards}
P.~Zhou, X.~Xie, Z.~Lin, and S.~Yan,
\newblock ``Towards understanding convergence and generalization of adamw,''
\newblock {\em IEEE TPAMI 2024}.

\bibitem{raymondaud2024probing}
Q.~Raymondaud, M.~Rouvier, and R.~Dufour,
\newblock ``Probing the information encoded in neural-based acoustic models of automatic speech recognition systems,''
\newblock {\em arXiv preprint arXiv:2402.19443}, 2024.

\end{thebibliography}

\end{document}